\documentclass[paper]{ieice}
\usepackage{graphicx,xcolor}
\usepackage[fleqn]{amsmath}
\usepackage{multirow}
\usepackage[psamsfonts]{amssymb}
\usepackage{amsxtra}
\usepackage{amsmath}
\usepackage{ulem}
\usepackage{soul}
\usepackage{threeparttable,comment}
\usepackage{cite,array}
\usepackage{subfig}
\usepackage{mathrsfs,color}
\usepackage{mathtools}

\usepackage{threeparttable,bm,amsmath,graphicx,amssymb,comment,array,multirow,threeparttable,cite,calligra,color}

\usepackage{comment}

\usepackage{amssymb}
\usepackage{amsfonts}

\usepackage{multirow}
\usepackage{cite}
\usepackage{textcomp}
\usepackage{amsmath,graphicx}

\makeatletter
\newcommand{\figcaption}[1]{\def\@captype{figure}\caption{#1}}
\newcommand{\tblcaption}[1]{\def\@captype{table}\caption{#1}}

\def\vector#1{\mbox{\boldmath $#1$}}

\makeatletter
\renewcommand\subsubsection{\@startsection{subsubsection}{3}{\z@}%
{0.5ex\@plus 0ex \@minus -.5ex}%
{0.5ex\@plus 0ex}
{\normalfont\normalsize\itshape}
}
\makeatother

\field{}
\title{Privacy-Preserving Support Vector Machine Computing Using Random Unitary Transformation} \authorlist{
\authorentry[maekawa-takahiro@ed.tmu.ac.jp]{Takahiro MAEKAWA}{n}{tmu}\MembershipNumber{}
\authorentry[kawamura-ayana@ed.tmu.ac.jp]{Ayana KAWAMURA}{n}{tmu}\MembershipNumber{}
\authorentry[nakachi.takayuki@lab.ntt.co.jp]{Takayuki NAKACHI}{m}{ntt}\MembershipNumber{}
\authorentry[kiya@tmu.ac.jp]{Hitoshi KIYA}{f}{tmu}\MembershipNumber{}}
\affiliate[tmu]{The authors are with Tokyo Metropolitan University, Hino-shi, 191-0065 Japan}
\affiliate[tmu]{The authors are with NTT Network Innovation Laboratories, Kanagawa, 239-0847 Japan}
\received{2015}{1}{1}
\revised{2015}{1}{1}
\begin{document}
\newcommand{\red}[1]{\textcolor{black}{#1}}
\maketitle
\begin{summary}
  A privacy-preserving support vector machine (SVM) computing scheme is proposed in this paper. Cloud computing  has been spreading in many fields. However, the cloud computing has some serious issues for end users, such as the unauthorized use of cloud services, data leaks, and privacy being compromised. Accordingly, we consider privacy-preserving SVM computing. We focus on protecting visual \red{information} of images by using a random unitary transformation. Some properties of the protected images are discussed. The proposed scheme enables us not only to protect images, but also to have the same performance as that of unprotected images even when using typical kernel functions such as the linear kernel, radial basis function(RBF) kernel and polynomial kernel. Moreover, it can be directly carried out by using well-known SVM algorithms, without preparing any algorithms specialized for secure SVM computing. In an experiment, the proposed scheme is applied to a face-based authentication algorithm with SVM classifiers to confirm the effectiveness.
\end{summary}
\begin{keywords}
  Support Vector Machine, Privacy-preserving,  random unitary transformation
\end{keywords}

\section{Introduction}
\label{sec:intro}
Cloud computing and edge computing have been spreading in many fields
with the development of cloud services. However, the computing
environment has some serious issues for end users, such as
the unauthorized use of cloud services, data leaks, and privacy being compromised due to
unreliabile providers and some accidents.
A lot of studies on secure, efficient, and flexible
communications, storage, and computation have been reported \cite{Nakamura2019,Ra2013,Huang, Lazzeretti, Barni,Lagendij}.
For securing data, full encryption with provable security (like RSA and 
AES) is the most secure option. However, many multimedia
applications have been seeking a trade-off in security to enable other
requirements, e.g., low processing demands, retaining bitstream
compliance, and flexible processing in the encrypted domain, so a
lot of perceptual encryption schemes have been studied to achieve a trade-off \cite{Lagendij, Ito1, Chuman2, Zhou, Kurihara_1, Kurihara, Chu_Kuri_1, Chu_Kuri_2, Chuman2018,Warit2019APSIPA}

In recent years, considerable efforts have been made in the
fields of  fully homomorphic encryption and multi-party
computation \cite{Araki1, Araki2, Lu, Toshinori}. However, these schemes can not be applied yet to SVM
algorithms, although it is possible to carry out some statistical
analysis of  categorical and ordinal data. Moreover, the schemes have
to prepare algorithms specialized for computing encrypted data.

Because of this, we propose a privacy-preserving SVM
computing scheme in this paper . We focus on images protected by
using
a random unitary transformation, which have been studied as one of
methods for cancelable biometrics \cite{Rathgeb, Nandakumar, Rane, Wright, Nakamura1, Nakamura2, Georghiades}, and then  consider some
properties of the protected images for secure SVM computing, where
images mean features extracted from data.
 As a result, the
proposed scheme enables us not only to protect images, but also to
have the same performance as that of unprotected images under some
useful kernel functions as isotropic stationary kernels.  Moreover, it can be directly carried out by
using well-known SVM algorithms, without preparing any algorithms
specialized for secure SVM computing. 
\red{SVM is a typical machine learning algorithm that allows us to use kernel tricks.
SVM is used as an example of machine learning algorithms based on the Euclidean distance or the inner product between vectors.
It is shown that the proposed scheme enables to maintain the Euclidean distance and the inner product, so the scheme can be also applied to other machine learning algorithms.}In an experiments, the proposed
scheme is applied to a face recognition algorithm with SVM classifiers
to confirm the effectiveness.

\section{Preparation}
\subsection{Support Vector Machine}
support vector machine (SVM) is a supervised machine learning algorithm that can be used for both classification and regression challenges, but it is mostly used in classification problems. In SVM, we input a feature vector $\vector{x}$ to the discriminant function as

\begin{equation}
\label{eq:eq_sign}
\begin{split}
y = \mathrm{sign}(\vector{\omega}^T\vector{x}+b)\\
\lefteqn{\hspace{-49mm}text{with}}\\
   \mathrm{sign}(u)=\begin{cases}
     1 & (u>1) \\
    -1 & (u\leq0)
  \end{cases},
  \end{split}
\end{equation}

where $\vector{\omega}$ is a weight parameter, and  $b$ is a bias.
SVM also has a technique called the "kernel trick", which is a function that takes low dimensional input space and transform it to a higher dimensional space. These functions are called kernels. The kernel trick can be applied to Eq. (\ref{eq:eq_sign}) to map an input vector on a further high-dimension feature space and then to linearly classify it on that space as
\begin{equation}
\label{eq:eq_kernel_sign}
  y = \mathrm{sign}(\vector{\omega}^T\phi(\vector{x})+b).
\end{equation}
The function $\phi(\vector{x}):\mathbb{R}^d\to\mathcal{F}$ maps an input vector $\vector{x}$ on high dimensional feature space $\mathcal{F}$, where $d$ is the number of the dimensions of features.
In this case, feature space $\mathcal{F}$ includes parameter $\vector{\omega}$ ($\vector{\omega}\in\mathcal{F}$).
The kernel function of two vectors $\vector{x}_i$, $\vector{x}_j$ is defined as
\begin{equation}
  K(\vector{x}_i,\vector{x}_j)=\langle \phi(\vector{x}_i), \phi(\vector{x}_j)\rangle,
\end{equation}
where $\langle\cdot, \cdot\rangle$ is an inner product. There are various kernel functions\cite{Genton}. For example, the radial basis function (RBF) kernel is given by
\begin{equation}
  \label{eq:rbf}
  K(\vector{x}_i,\vector{x}_j)=\exp(-\varUpsilon \| \vector{x}_i - \vector{x}_j \|^2 )
\end{equation}
and the polynomial kernel is provided by
\begin{equation}
  K(\vector{x}_i,\vector{x}_j)=(1+\vector{x}_i^T\vector{x}_j)^l,
\end{equation}
where $\varUpsilon$ is a high parameter for deciding the complexity of boundary determination, $l$ is a parameter for deciding the degree of the polynomial, and $T$ indicates a transpose.

\red{This paper aims to propose a new framework to carry out some machine learning algorithms with protected vectors.
SVM is used to demonstrate the effectiveness of the proposed scheme as one of machine learning algorithms.}

\subsection{Scenario}
\begin{figure}[t]
  \centering\includegraphics[width = 8cm]{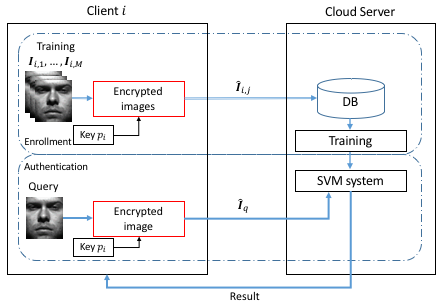}
  \caption{Scenario}
  \label{fig:architecture}
\end{figure}

Figure \ref{fig:architecture} illustrates the scenario used in this paper. In the enrollment task, client $i$, $i \in \{1,2,...,N\}$, prepares training images $ \boldmath{I}_{i,j}, j \in \{1,2,...,M\}$. Next the client creates protected images $\hat{\boldmath{I}}_{i,j}$ by using a secret key $p_i$ and sends them to a cloud server. The server stores them and implements learning with the protected images for a classification problem.

In the authentication task, client $i$ creates a protected image as a query and sends it to the server. The server carries out a classification problem with a learning model prepared in advance, and then returns the result to client $i$.

Note that the cloud server has no secret keys and the classification problem can be directly carried out by using a well-known SVM algorithm.  In the other words, the server does not have to prepare any algorithms specialized for the classification in the encrypted domain.

\section{Proposed framework}
In this section, protected images generated by using a random unitary matrix are conducted, and a SVM computation scheme with the protected images is proposed under the use of some kernel functions.
\subsection{Protection of visual \red{information}}
\label{sec:Uniprotect}
Protection schemes of visual information based on unitary transformations have been studied as one method for cancelable biometrics\cite{Rathgeb, Wright, Nakamura1,Nakamura2, Nandakumar, Rane}. This paper has been inspired by those studies.

Let us transform an image $\boldmath{I}_{i,j}$ with $X \times Y$pixels into a vector $\vector{\mathrm{f}_{i,j}} =\{\red{l}_{i,j}(0),...,\red{l}_{i,j}(d-1)\}^T \in \mathbb{R}^d, d = X \times Y$, where $\red{l}_{i,j}(k), k = 1,2,...,d-1$ is a pixel value of $\boldmath{I}_{i,j}$.
A vector $\vector{\mathrm{f}_{i,j}} \in \mathbb{R}^d $ is protected by a unitary matrix having randomness with a key $p_i$, $ \vector{\mathrm{Q}_{p_i}} \in \red{\mathbb{C}^{d \times d}}$ as
\begin{equation}
  \label{eq:trans}
  \hat{\vector{\mathrm{f}}}_{i,j}=T(\vector{\mathrm{f}}_{i,j},{p_i})=\vector{\mathrm{Q}}_{p_i}\vector{\mathrm{f}}_{i,j},
\end{equation}
where $\hat{\vector{\mathrm{f}}}_{i,j}$ is a protected vector. Various generation schemes of $\vector{\mathrm{Q}}_{p_i}$ have been studied to design unitary or orthogonal random matrices such as Gram-Schmidt-based methods, random permutation matrices and random phase matrices\cite{Nakamura2, Nakamura1}. For example, the Gram-Schmidt-based methods are applied to a pseudo-random matrix to generate $\vector{\mathrm{Q}}_{p_i}$. Security analysis of the protection schemes have been also considered in terms of brute-force attacks, diversity and irreversibility.

\subsection{SVM with protected images}
  \label{sec:SVMpro}
\subsubsection{Properties}
Protected images generated according to Eq. (\ref{eq:trans}) have the following properties under $p_i=p_s$\cite{Nakamura2}.

\noindent
\hspace{15pt}Property 1 : Conservation of Euclidean distances:\\
          \begin{equation}\| \vector{\mathrm{f}}_{i,j} - \vector{\mathrm{f}}_{s,t} \|^2  =  \| \hat{\vector{\mathrm{f}}}_{i,j} - \hat{\vector{\mathrm{f}}}_{s,t} \|^2. \end{equation} \\
\hspace{15pt}Property 2 : Conservation of inner products:\\
          \begin{equation}\langle \vector{\mathrm{f}}_{i,j},\vector{\mathrm{f}}_{s,t}\rangle=\langle\hat{\vector{\mathrm{f}}}_{i,j},\hat{\vector{\mathrm{f}}}_{s,t}\rangle, \end{equation}\\
\hspace{15pt}Property 3 : Conservation of correlation coefficients: \\
           \begin{equation}
             \frac{\langle \vector{\mathrm{f}}_{i,j},\vector{\mathrm{f}}_{s,t}\rangle}{\sqrt{\langle \vector{\mathrm{f}}_{i,j},\vector{\mathrm{f}}_{s,t}\rangle}\sqrt{\langle \vector{\mathrm{f}}_{i,j},\vector{\mathrm{f}}_{s,t}\rangle}}=\frac{\langle\hat{\vector{\mathrm{f}}}_{i,j},\hat{\vector{\mathrm{f}}}_{s,t}\rangle}{\sqrt{\langle\hat{\vector{\mathrm{f}}}_{i,j},\hat{\vector{\mathrm{f}}}_{s,t}\rangle}\sqrt{\langle\hat{\vector{\mathrm{f}}}_{i,j},\hat{\vector{\mathrm{f}}}_{s,t}\rangle}}.
             \end{equation}\\
where $\vector{\mathrm{f}}_{s,t}$ is a vector of another client $s, s \in \{1,2,...,N\}$, who has M training samples $\mathrm{g}_{s,t}, t \in \{1,2,...,M\}$.

\subsubsection{Classes of kernels}
We consider applying encrypted images to the kernel trick. In the
case of using the RBF kernel, the following relation is satisfied from
property 1 and Eq.(4):
\begin{eqnarray}
  \label{eq:propKernel}
  K(\hat{\vector{\mathrm{f}}}_{i,j},\hat{\vector{\mathrm{f}}}_{s,t})
  &=&\exp(- \varUpsilon \| \hat{\vector{\mathrm{f}}}_{i,j} - \hat{\vector{\mathrm{f}}}_{s,t} \|^2 )\nonumber\\
  &=&K(\vector{\mathrm{f}}_{i,j},\vector{\mathrm{f}}_{s,t}).
\end{eqnarray}
Therefore, protected images do not have any influence when using kernel functions based on Euclidean distance, such as the RBF kernel.We call the class of these Euclidean distance based kernel functions class 1 in this paper. 

In addition, from property 2, we can also use a kernel that depends only on the inner products between two vectors.The polynomial kernel and linear kernel are in this class, referred to as class 2.
Therefore, following relations are satisfied, under property 2,
\begin{eqnarray}
K(\hat{\vector{\mathrm{f}}}_{i,j},\hat{\vector{\mathrm{f}}}_{s,t})&=&\langle\hat{\vector{\mathrm{f}}}_{i,j},\hat{\vector{\mathrm{f}}}_{s,t}\rangle \nonumber \\
&=&K({\vector{\mathrm{f}}}_{i,j},{\vector{\mathrm{f}}}_{s,t})
\end{eqnarray}
\begin{eqnarray}
K(\hat{\vector{\mathrm{f}}}_{i,j},\hat{\vector{\mathrm{f}}}_{s,t})&=&(1+\langle\hat{\vector{\mathrm{f}}}_{i,j},\hat{\vector{\mathrm{f}}}_{s,t}\rangle)^{l} \nonumber \\
&=&K({\vector{\mathrm{f}}}_{i,j},{\vector{\mathrm{f}}}_{s,t}).
\end{eqnarray}

\subsubsection{Dual problem}
Next, we consider binary classification that is the task of classifying the elements of a given set. A dual problem for implementing a SVM classifier with protected images is expressed as
\footnotesize
\begin{equation}
  \label{eq:eq_dual_kernel}
  \begin{split}
    &\max_\alpha\ \left(-\frac{1}{2}\sum_{\substack{i,s \in N\\j,t \in M}}\alpha_{i,j} \alpha_{s,t} y_{i,j} y_{s,t} \langle \phi(\hat{\vector{\mathrm{f}}}_{i,j}, \hat{\vector{\mathrm{f}}}_{s,t})\rangle + \sum_{\substack{i \in N\\j \in M}}\alpha_{i,j}
    \right)\\
    &s.t.\ \sum_{\substack{i \in N \\ j \in M}}\alpha_{i,j} y_{i,j} = 0, 0\leq\alpha_{i,j}\leq C,
  \end{split}
\end{equation}
\normalsize
where $y_{i,j}$ and $y_{s,t}$$\in\{+1,-1\}$ are correct labels for each piece of training data, $\alpha_{i,j}$ and $\alpha_{s,t}$ are dual variables and C is a regular coefficient.
If we use kernel class 1 or class 2 described above, the inner product $\langle\phi(\hat{\vector{\mathrm{f}}}_{i,j}), \phi(\hat{\vector{\mathrm{f}}}_{s,t})\rangle$ is equal to $K(\vector{\mathrm{f}}_{i,j},\vector{\mathrm{f}}_{s,t})$.
Therefore,even in the case of using protected images, the dual problem with protected images is reduced to the same problem as that of the original images. This conclusion means that the use of the proposed images gives no effect to the performance of the SVM classifier under kernel class 1 and class 2.

\subsection{Relation among keys}
\label{sec:key}
As shown in Fig \ref{fig:architecture}, a protected image $\hat{\boldmath{I}}_{i,j}$ is generated from training image $\boldmath{I}_{i,j}$ by using a key $p_i$. Two relations among keys are summarized here.
\subsubsection{Key condition 1: $p_1=p_2=...=p_N$}
The first key choice is to use a common key for all clients, namely, $p_1=p_2=...=p_N$. In this case, all protected images satisfy the properties described in \ref{sec:SVMpro}, so the SVM classifier has the same performance as that of using the original images.
\subsubsection{Key condition 2: $p_1 \neq p_2 \neq .. .\neq p_N$}
The second key choice is to use a different key for each client, namely $p_1 \neq p_2 \neq .. .\neq p_N$. In this case, the three properties are satisfied only among images with a common key. This key condition allows us to enhance the robustness of security against various attacks as discussed later.\\

Under this key condition 2, we can consider two type spoofing attacks. Fist one is the case that secred keys $p_1 \neq p_2 \neq .. .\neq p_N$ leak out and an attacker use them. The attacker can try to authorize the system with the leaked key. And another is case that a original images of some client leak out. In this case, the attacker can authorize with the original images by transformed by some key which is created by the attacker as discribed in our experiments.

\section{Experimental Results}
The propose scheme was applied to facial recognition experiments that were carried out as a dual problem.
\subsection{Data Set}
\begin{figure}[t]
\centering
  \begin{tabular}{c c c c}
     \begin{minipage}[b]{0.15\hsize}
 	     \centering\includegraphics[width = 1.5cm]{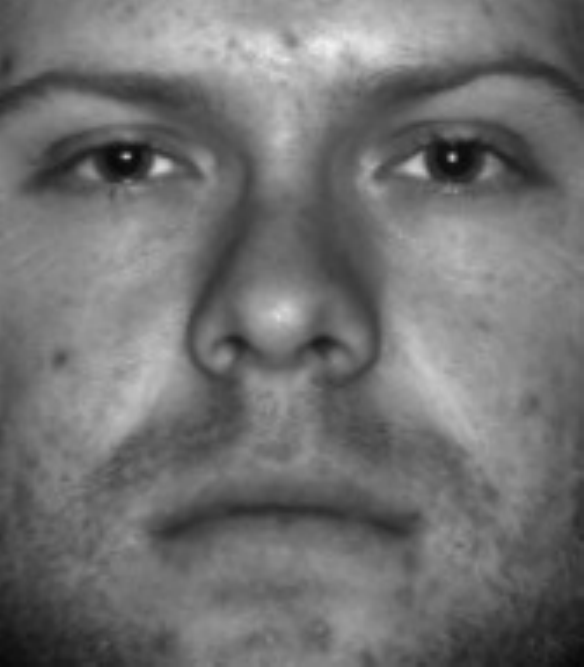}
     \end{minipage}
      &
     \begin{minipage}[b]{0.15\hsize}
        \centering\includegraphics[width = 1.5cm]{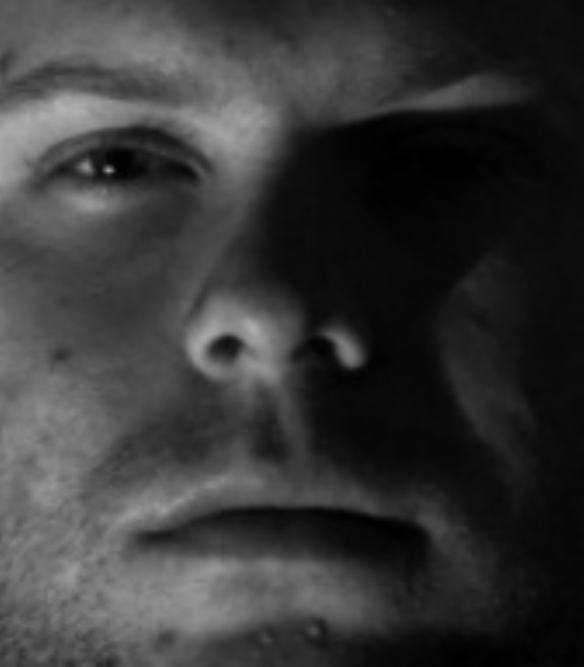}
    \end{minipage}
      &
    \begin{minipage}[b]{0.15\hsize}
   	 \centering\includegraphics[width = 1.5cm]{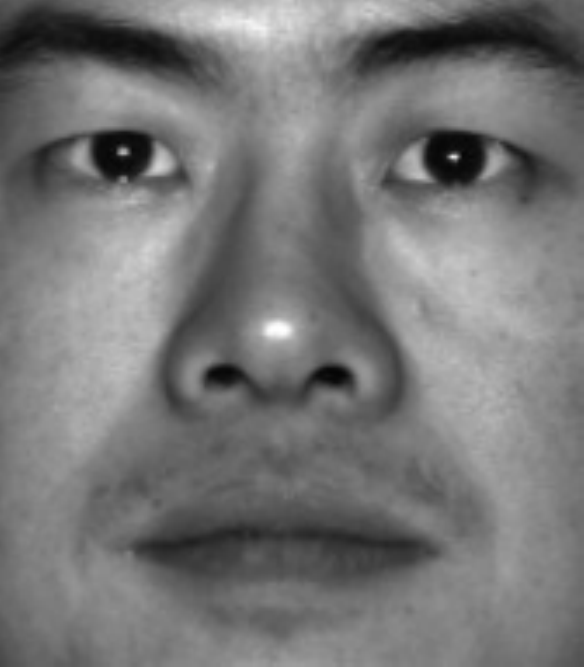}
    \end{minipage}
      &
    \begin{minipage}[b]{0.15\hsize}
      \centering\includegraphics[width = 1.5cm]{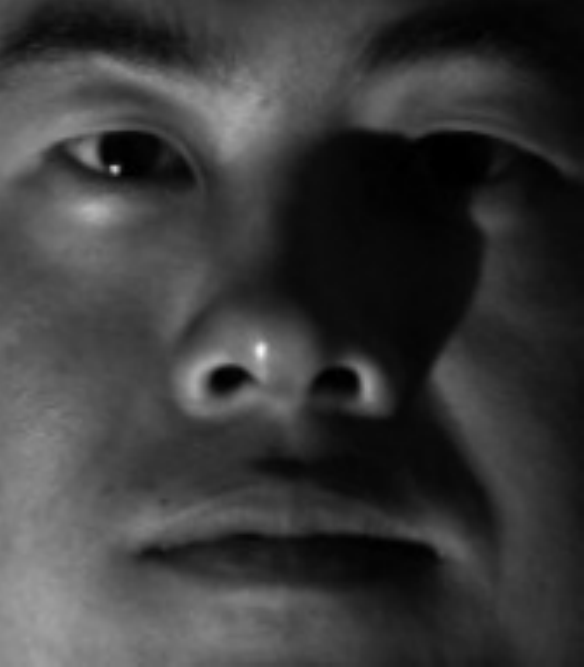}
    \end{minipage}
  \\
    \multicolumn{2}{c}{(a) person1}
     &
    \multicolumn{2}{c}{(b) person2}
  \end{tabular}
\caption{Examples of Extended Yale Face Database B}
\label{fig:db}
\end{figure}

\begin{figure}[t]
\centering
  \begin{tabular}{c c}
  \begin{minipage}[b]{0.2\hsize}
      \begin{center}
 	 \centering\includegraphics[width = 2cm]{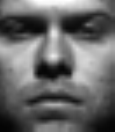}
      \end{center}
    \end{minipage}
    &
     \begin{minipage}[b]{0.2\hsize}
      \begin{center}
        \centering\includegraphics[width = 2cm]{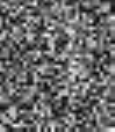}
      \end{center}
    \end{minipage}
  \\
   (a) image
   &
   (b) protected
\end{tabular}
\caption{An example of protection}
\label{fig:imageProtected}
\end{figure}

We used the Extended Yale Face Database B\cite{Georghiades}, which consists of $38 \times 64=2432$ frontal facial images with $192\times168$-pixels for $N=38$ people like Fig. \ref{fig:db}.
\red{It is assumed that there were clients (users), a cloud server, and an attaker (a heinous third party) in this paper. 36 people were used as clients and 64 images for each person were divided into half randomly for training data samples and queries.
1 person was used as an attacker from the database and 32 images of the attacker were used as queries.}
 We used random permutation matrices as \red{an example of} unitary matrices to produce protected images\red{, although there are other transformations such as the Gram-Schmidt-based method.} 
\red{It is known that random permutation matrices have an advantage in terms of less computational complexity compared with the Gram-Schmidt-based method\cite{Nakamura2}.
Any unitary transformations with randomness are applicable to the proposed scheme.}
Besides, the RBF kernel and linear kernel were used, where they belong to kernel class 1 and class 2, respectively.
The protection was applied to images with 1216 dimensions generated by the down-sampling method\cite{Wright}. The down-sampling method divides an image into non-overlapped blocks and then calculates the mean value in each block. Figure \ref{fig:imageProtected} shows the examples of an original image and the protected one. Here, the protected image was created by a random permutation matrix which consists of 0 and 1.

\subsection{Results and Discussion}
In facial recognition with SVM classifiers, one classifier is created for each enrollee. The classifier outputs a predicted class label and a classification score for each query image $\hat{\boldmath{I}}_{q}$, where $\hat{\boldmath{I}}_{q}$ is a protected image generated from the image of a query $\boldmath{I}_{q}$. The classification score is the distance from a query to the boundary range. The relation between the classification score $S_q$ and a threshold $\tau$ for a positive label of $\boldmath{I}_{q}$ is given as
\begin{equation}
if \ S_q \geq \tau \ then\  accept;\  else\ reject.
\label{accept}
\end{equation}
In the experiment, the false reject rate(FRR), false accept rate(FAR), and equal error rate(EER) at which FAR is equal to FRR, were used to evaluate the performance.
\red{
As described in 4.1, face images of 37 people including 1 attacker were prepared and there were 64 images for each person.
$36 \times 32=1152$ images of 36 people were used for training, and other 1152 images of 36 people and 32 images of the attacker were used as query ones for authentication respectively, under various key conditions.
}
\subsubsection{$p_1=p_2=...=p_N$}\par
\begin{figure}[t]
  \captionsetup[subfigure]{justification=centering}
\centering
\subfloat[Linear kernel ($C=1$)\label{fig:fig1}]
{\includegraphics[width=7.5cm]{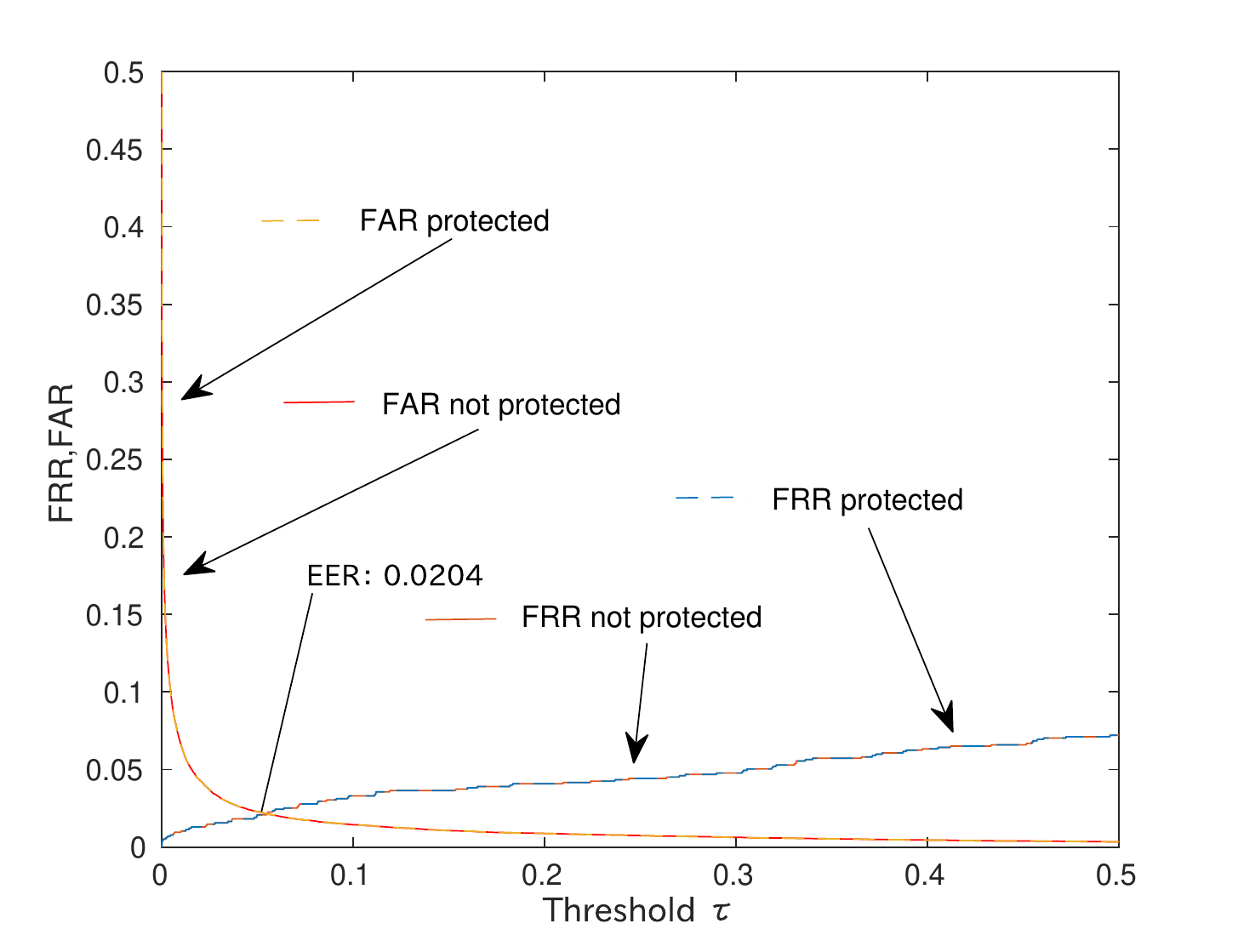}}

\centering

\centering
\subfloat[RBF kernel ($C=34$, $\varUpsilon=81$)\label{fig:fig1}]
{\includegraphics[width=7.5cm]{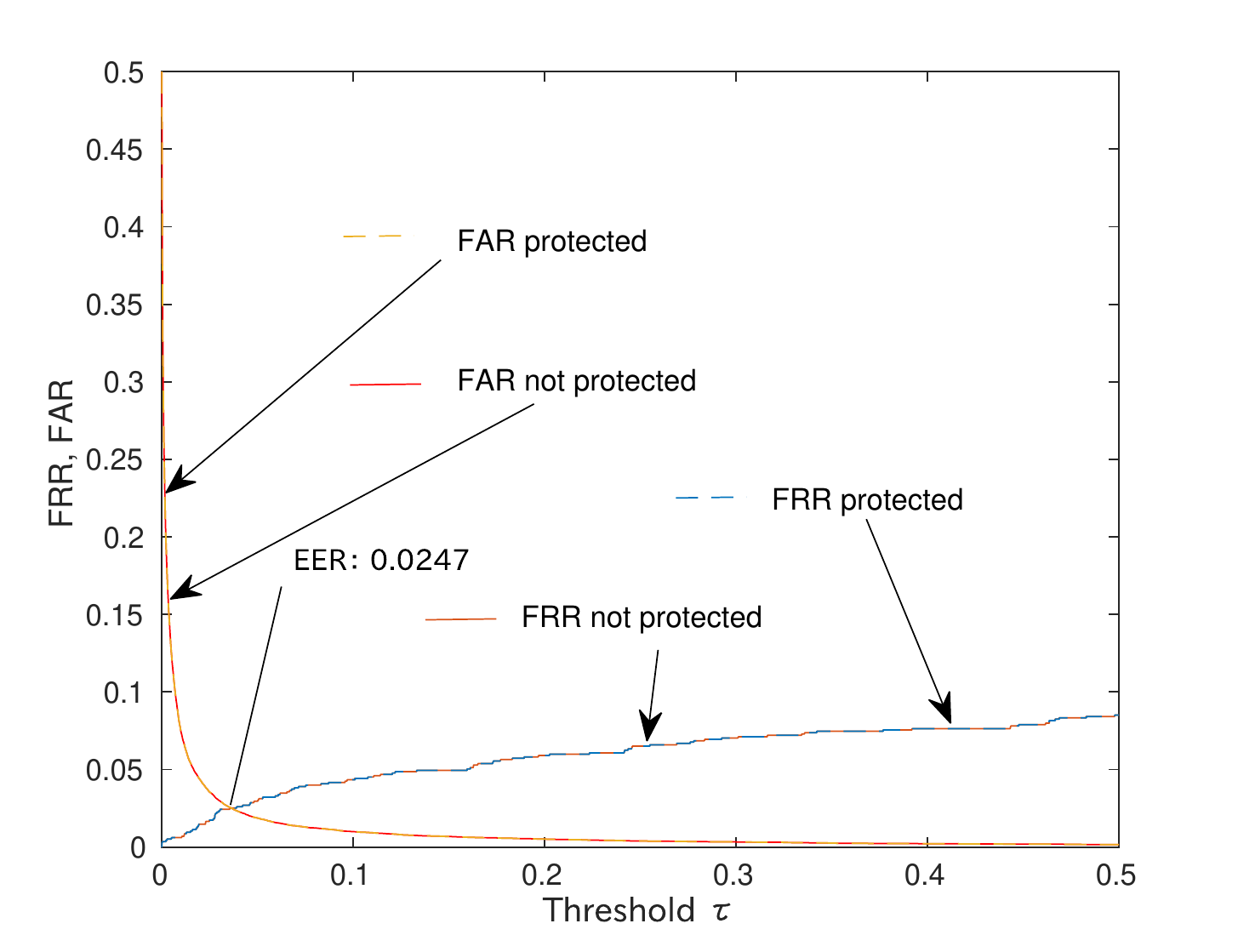}}
\caption{FAR and \red{FRR} ($p_1=p_2=...=p_N$)}\label{fig:result_1}
\end{figure}

Figure \ref{fig:result_1} shows results in the case of using key condition 1. The results demonstrate that SVM classifiers with protected images (protected in Fig. \ref{fig:result_1}) performed the same as SVM classifiers with the original images (not protected in Fig. \ref{fig:result_1}). 

\red{In the experiment, when 32 images of person 1 were used as query ones, the FRR value of person 1, $FRR_1$, under a $\tau$ value was calculated as follows. 
The number of images $r_1$, which were rejected as another person from Eq.(\ref{accept}), was calculated, and then the rate of the rejected images was calculated as $FRR_1 = r_1 / 32$.
Finally, the average of $FRR_i$ values over 36 people was obtained as $FRR = \sum_{i=1}^{36}(FRR_i / 36)$.}

\red{
The FAR value of person 1, $FAR_1$, under a $\tau$ value was calculated as follows. When $35 \times 32$ images without images of person 1 were used as query ones, the number of images $s_1$, which were accepted as person 1 from Eq.(\ref{accept}), was calculated, and then the rate of the accepted images was calculated as $FAR_1 = s_1 / (35 \times 32)$. 
Finally, the average of $FAR_i$ values over 36 people was obtained as $FAR = \sum_{i=1}^{36}(FAR_i / 36)$.
}
From the results, it is confirmed that the proposed scheme gives no effect to the performance of SVM classifiers under key condition 1.
\subsubsection{$p_1 \neq p_2 \neq .. .\neq p_N$}\par
\begin{figure}[t]
  \centering\includegraphics[width = 7.5cm]{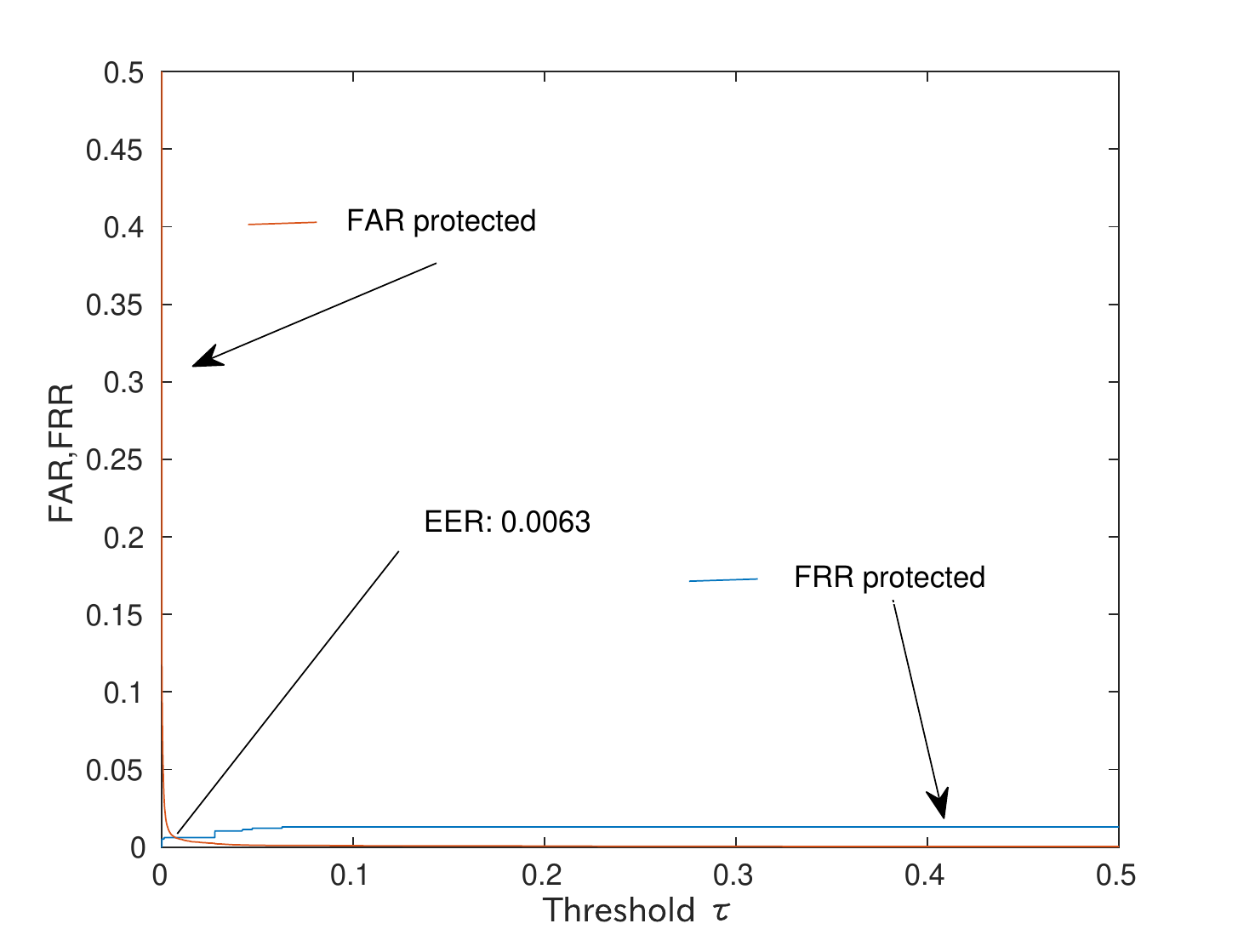}
 \caption{FAR and \red{FRR} (RBF kernel, $p_1 \neq p_2 \neq .. .\neq p_N$)}
 \label{fig:result_2}
\end{figure}

Figure \ref{fig:result_2} shows results in the case of using key condition 2. In this condition, it is expected that a query will be authenticated only when it meets two requirements, i.e. the same key and the same person, although only the same person is required  under key condition1. Therefore, the performances in Fig. \ref{fig:result_2} were slightly different from those in Fig. \ref{fig:result_1}, so the FAR performances for key condition 2 were better due to the strict requirements.

\subsubsection{Unauthorized outflow}\par
\red{Next, it is assumed that a key or an image leaks from a client.
An attacker (a heinous third party) may be able to spoofs user $i$ with a leaked key or leaked images of person $i$.
If an private image leaks, the visual information can not be protected, but the third party's spoofing attack may be able to be protected by using encrypted images.
In this experiment, we evaluated FAR performances when a key or images leaked out and an attacker spoofed person 1 to 36 with the leaked key or the leaked images. 
When a key leaked, the attacker spoofed a user with the leaked key and images of the attcker.
In contrast, when images leaked, the attacker spoofs a user with the leaked images and a key prepared by the attacker.
Protected images to spoof a user were generated by the leaked images and the key. 
When 32 images prepared by the attacker were used as query ones, the FAR value of user 1 was calculated as follows. The number of images $s_i$, which were accepted as person 1 from Eq.(\ref{accept}) was calculated, and then the rate of the false accepted images was calculated as $FAR_1 = s_1 / 32$. Finally, the average of $FAR_i$ values over 36 users was obtained as $FAR = \sum_{i=1}^{36}(FAR_i / 36)$.
}

\begin{figure}[t]
  \centering\includegraphics[width = 7.5cm]{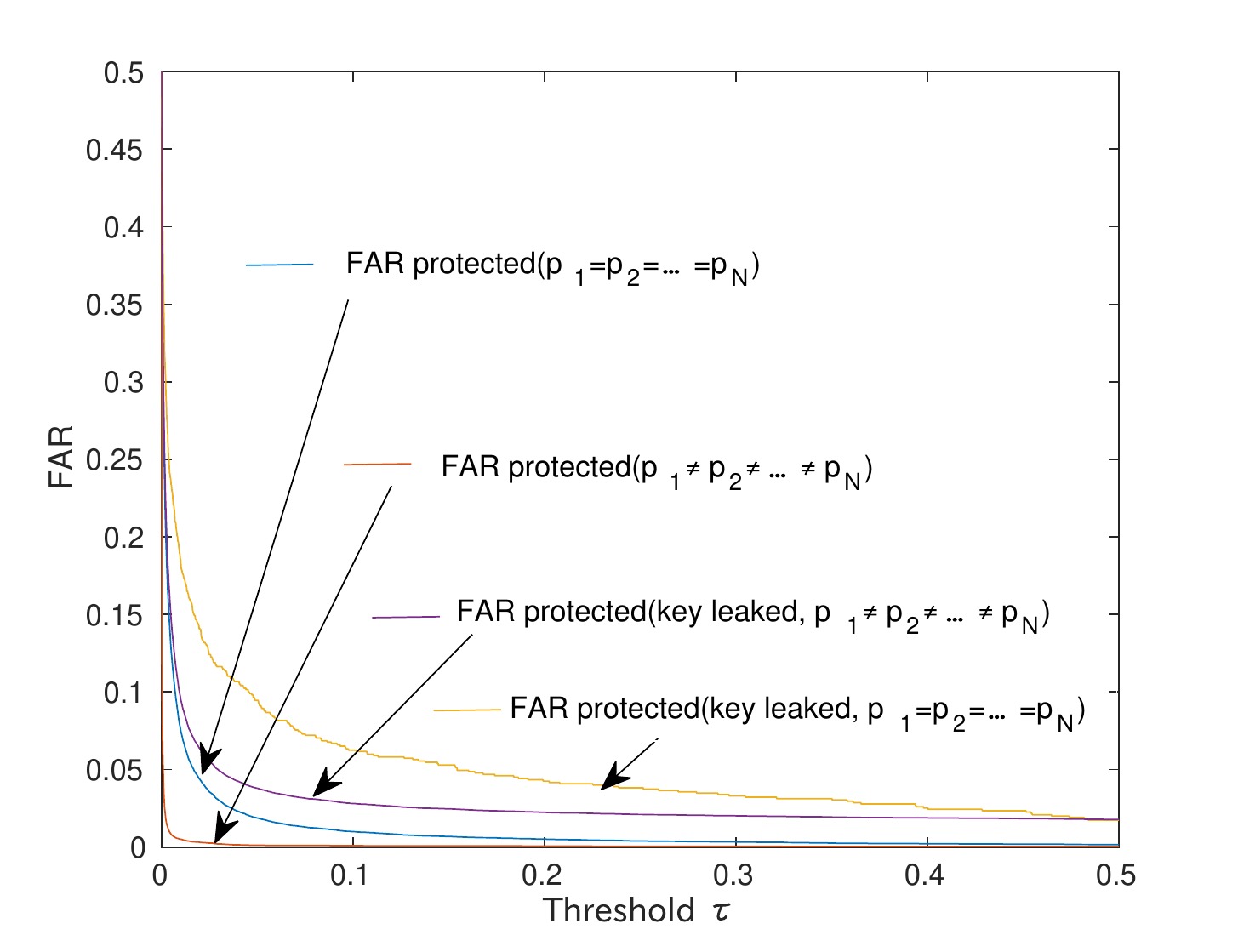}
 \caption{FAR with leaked keys (RBF kernel)}
 \label{fig:key_leak}
\end{figure}

Figure \ref{fig:key_leak} shows the FAR performance in the case that a key $p_i$ leaked out. In this situation, \red{the attacker} could use the key $p_i$ without any authorization as spoofing attacks. 
\red{
"FAR protected (key leaked, $p_1 = p_2 = ...= p_N$)" indicates FAR values when clients used a same key and the key leaked out. "FAR protected (key leaked, $p_1 \neq p_2 \neq ... \neq p_N$)" indicates FAR values when each client used a different key and the key leaked out.
"FAR protected (key leaked, $p_1 \neq p_2 \neq ... \neq p_N$)" was better than "FAR protected (key leaked, $p_1 = p_2 = ...= p_N$)".
Therefore, it is confirmed that the security against the spoof with the leaked key is enhanced, if we can use key condition 2.}

\begin{figure}[t]
  \centering\includegraphics[width = 7.5cm]{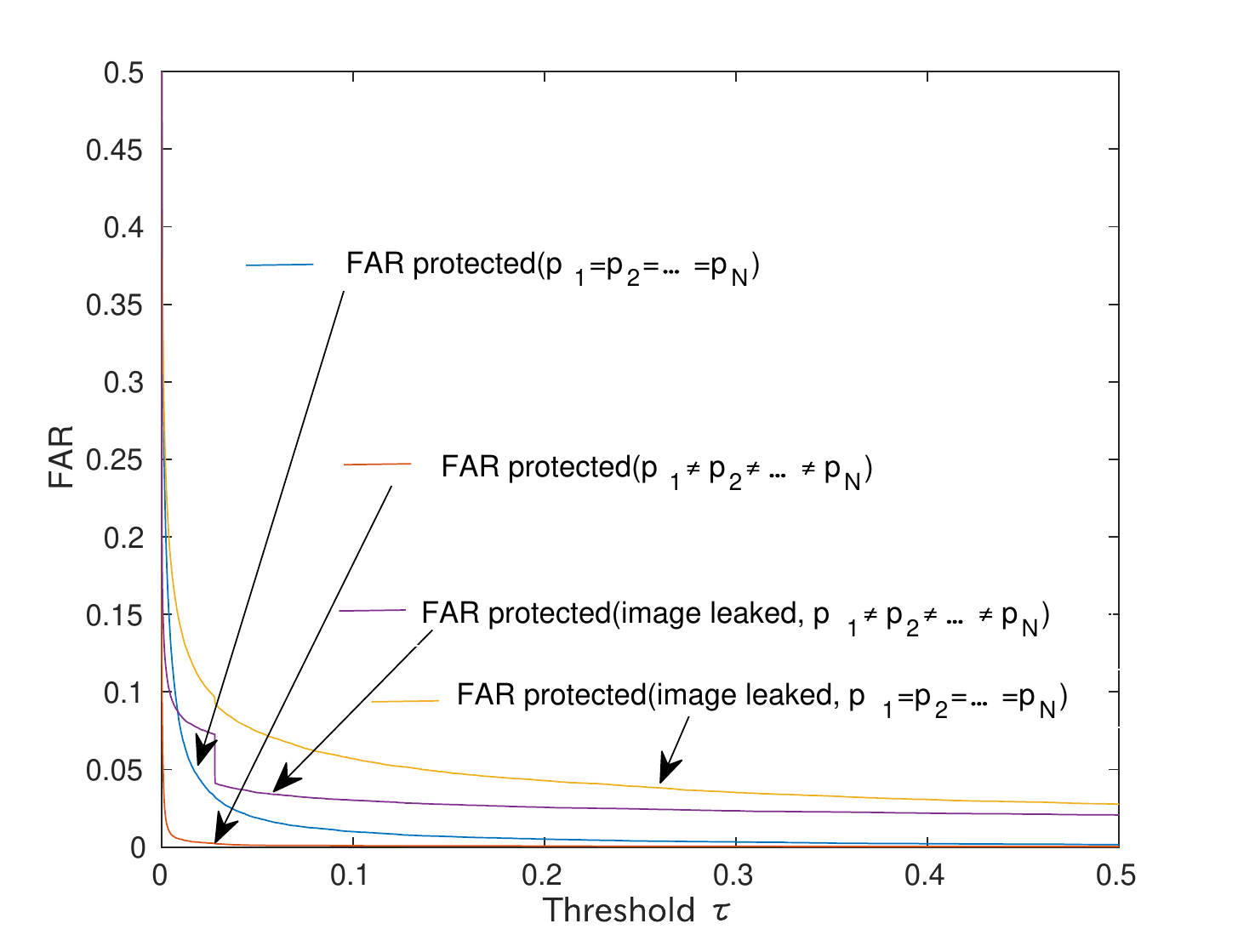}
 \caption{FAR with leaked original images (RBF kernel)}
 \label{fig:image_leak}
\end{figure}

Figure \ref{fig:image_leak} is the FAR performance in the case that images of person $i$ leaked out. 
\red{"FAR protected (image leaked, $p_1 = p_2 = ...= p_N$)" indicates FAR values when clients used a same key and images of person $i$ leaked out. "FAR protected (image leaked, $p_1 \neq p_2 \neq ... \neq p_N$)" indicates FAR values when each client used a different key and the images of person $i$ leaked out.
As well as Fig. \ref{fig:key_leak}, FAR values under the use of different keys were lower than FAR ones under the same key.}

From these results, \red{FAR values under the use of different keys are improved not only when authentication is carried out by an enrolled user, but also when an attacker spoofs users using a leaked key or a leaked image.}
Therefore, the use of key condition 2 enhances the robustness of the security against spoofing attacks.

\section{conclusion}
In this paper, we proposed a privacy-preserving SVM computing scheme with
protected images. It was shown that images protected by a
unitary transform has some useful properties,  and  the properties
allow us to securely compute SVM algorithms without any degradation of
the performances. Besides, two key conditions were considered to
enhance the robustness of the security against various attacks.  Some
face-based authentication experiments using SVM classifiers were
also demonstrated to experimentally confirm the effectiveness of the
proposed scheme.

\subsection*{Acknowledgements}
This work was partially supported by Grant-in-Aid for Scientific
Research(B), No.17H03267, from the Japan Society
for the Promotion Science.


\profile[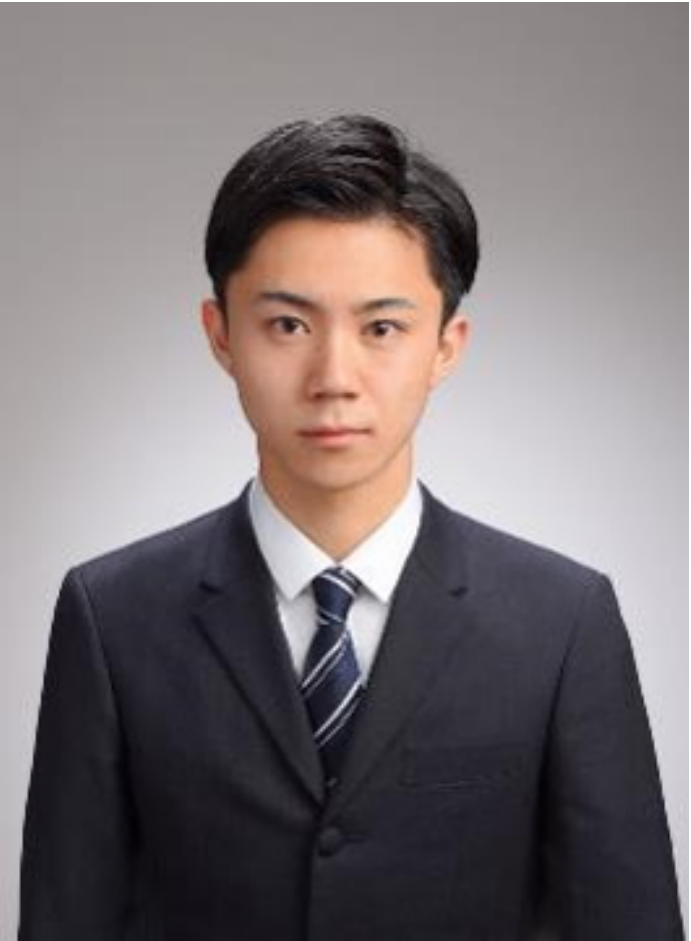]{Takahiro Maekawa}{
received his B.Eng. degree
from Tokyo Metropolitan University, Japan in
2017. He graduated a Master course at Tokyo Metropolitan University in 2019. His
research interests are in the area of image processing.}
\profile[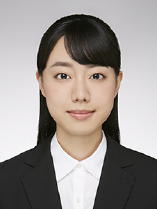]{Ayana Kawamura}{
received her B.Eng. degree
from Tokyo Metropolitan University, Japan in
2018. Since 2018, she has been a Master course
student at Tokyo Metropolitan University. Her
research interests are in the area of image processing.}
\profile[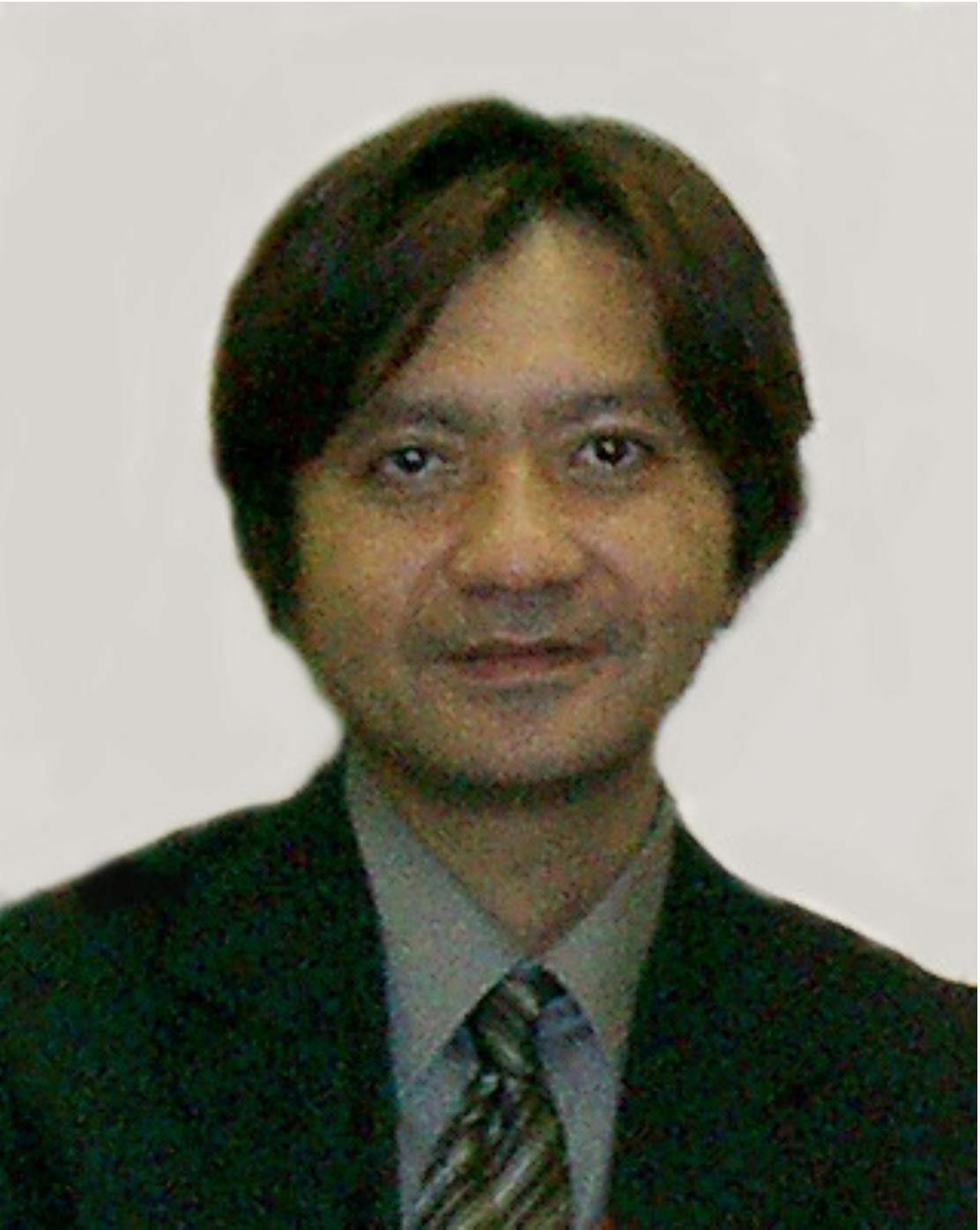]{Takayuki Nakachi}{received the Ph.D. degree in electrical engineering from Keio University, Tokyo, Japan, in 1997. Since he joined Nippon Telegraph and Telephone Corporation (NTT) in 1997, he has been engaged in research on super-high-definition image/video coding, media transport technologies. From 2006 to 2007, he was a visiting scientist at Stanford University. He also actively participates in MPEG international standardization. His current research interests include communication science, information theory and signal processing.
He received the 26th TELECOM System Technology Award, the 6th Paper Award of Journal of Signal Processing and the Best Paper Award of IEEE ISPACS2015. Dr. Nakachi is a member of the Institute of Electrical and Electronics Engineers the Institute of Electronics (IEEE) and the Information and Communication Engineers (IEICE) of Japan.}
\profile[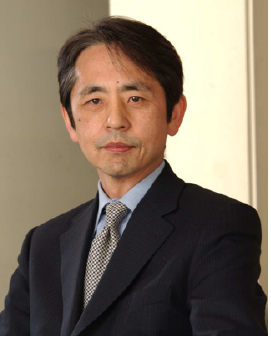]{Hitoshi Kiya}{received his B.E and M.E. degrees from Nagaoka University of Technology, in 1980 and 1982 respectively, and his Dr. Eng. degree from Tokyo Metropolitan University in 1987. In 1982, he joined Tokyo Metropolitan University, where he became a Full Professor in 2000. From 1995 to 1996, he attended the University of Sydney, Australia as a Visiting Fellow. He is a Fellow of IEEE, IEICE and ITE. He currently serves as President-Elect of APSIPA, and he served as Inaugural Vice President (Technical Activities) of APSIPA from 2009 to 2013, and as Regional Director-at-Large for Region 10 of the IEEE Signal Processing Society from 2016 to 2017. He was also President of the IEICE Engineering Sciences Society from 2011 to 2012, and he served there as a Vice President and Editor-in-Chief for IEICE Society Magazine and Society Publications. He was Editorial Board Member of eight journals, including IEEE Trans. on Signal Processing, Image Processing, and Information Forensics and Security, Chair of two technical committees and Member of nine technical committees including APSIPA Image, Video, and Multimedia Technical Committee (TC), and IEEE Information Forensics and Security TC. He has organized a lot of international conferences, in such roles as TPC Chair of IEEE ICASSP 2012 and as General Co-Chair of IEEE ISCAS 2019. He has received numerous awards, including six best paper awards.}
\end{document}